\documentclass[aps,prl,twocolumn,showpacs,groupedaddress]{revtex4}
\usepackage{amsfonts}
\usepackage{amssymb}
\usepackage{amsmath}
\usepackage{graphicx}
\usepackage{epstopdf}
\usepackage{dcolumn}   % column centering
\usepackage{bm}        % bold math
\usepackage{flafter}

\begin{document}

%\draft

\title{Spontaneous stoichiometry change in a single crystal of  $(Ba_{1-x}K_x)Fe_2As_2 $ produced  by rapid heating method}

\date{\today}

\author{P. Reuvekamp}
\affiliation{Department of Physics, Brock University,
St. Catharines, Ontario, L2S 3A1, Canada}
\author{F. S. Razavi}
\affiliation{Department of Physics, Brock University,
St. Catharines, Ontario, L2S 3A1, Canada}
\author{C. Hoch}
\affiliation{Max-Planck-Institut f\"{u}r Festk\"{o}rperforschung,
Heisenbergstra$\rm\beta$e 1, D-70569 Stuttgart, Germany}
\author{J. S. Kim}
\affiliation{Max-Planck-Institut f\"{u}r Festk\"{o}rperforschung,
Heisenbergstra$\rm\beta$e 1, D-70569 Stuttgart, Germany}
\author{R. K. Kremer}
\affiliation{Max-Planck-Institut f\"{u}r Festk\"{o}rperforschung,
Heisenbergstra$\rm\beta$e 1, D-70569 Stuttgart, Germany}
\author{A. Simon}
\affiliation{Max-Planck-Institut f\"{u}r Festk\"{o}rperforschung,
Heisenbergstra$\rm\beta$e 1, D-70569 Stuttgart, Germany}

\begin{abstract}

To prevent the loss of K in  growing single crystals of Ba$_{1-x}$K$_x$Fe$_2$As$_2$ we developed a rapid heating Sn-flux  method. Large single crystals with the optimal superconducting transition temperature T$_C \approx 38 $K were obtained and their structural, chemical and superconducting properties were investigated. Additionally, the effect of post growth  annealing at different temperatures were also examined on these crystals. Scanning electron microscopy microprobe studies on a samples with the composition goal of Ba$_{25}$K$_{75}$Fe$_2$As$_2$ revealed a sharp separation of two
phases with compositions that are suggestive rational ratios of the K and Ba content.

\end{abstract}
\smallskip


\maketitle
The finding of high-$T_C$ superconductivity up to 55 K  in the layered FeAs-based compounds and the observation of  a spin-density wave transition (SDW) at $\sim$ 140 K \cite{kamihara,takahashi,rotter} has stimulated broad  research to investigate the nature of the superconducting phase and the role of the SDW for superconductivity in these systems.
Electronic and magnetic properties reported so far, indicate some
similarities between the properties of the FeAS-based systems and those of high-$T_C$ oxocuprates.
In order to study these relationships more deeply and to refine some of the published preliminary  results,  growth of large  single-crystals with well defined and reproducible stoichiometry is crucial. Similar to high-$T_C$ oxocuprates the growth of homogenous, and impurity free single crystal of the FeAS-based compounds were found to be difficult. So far, rather small single-crystal of the layered oxopnictides LnFeAs(O$_{1-x}$F$_x$) (Ln=Pr, Nd,
Sm) were grown by using a NaCl/KCl flux method\cite{hashimoto,zhigadlo}.  Larger crystals of the AE$_{1-x}$K$_x$Fe$_2$As$_2$ (AE = Ba, Sr) were obtained by using either a Sn flux or a  self flux method.\cite{ni,chen1,ronning,Luo} Although both methods provide single-crystals  with nominally the same structure and lattice parameters as those  of powder samples, partly significant  differences in the electronic and magnetic properties between single-crystals and polycrystalline samples were observed.

At room temperature BaFe$_2$As$_2$ crystalizes with the well-known  tetragonal ThCr$_2$Si$_2$ structure-type (space group $I{\rm 4}mmm$.\cite{rotter} At $\sim$140 K polycrystalline samples show a first-order phase transition  to an orthorhombic structure (space group $Fmmm$) accompanied by a magnetic phase transition.\cite{rotter} The crystal structure of BaFe$_2$As$_2$ can be described as alternate stacks along the tetragonal axis of Ba and Fe$_2$As$_2$ layers.
Partial substitution of Ba$^{2+}$ by monovalent K or Cs induces superconductivity with
$T_C$'s  as high as 38 K\cite{rotter,sasmal,chen2}. First studies of polycrystalline samples of Ba$_{1-x}$K$_x$Fe$_2$As$_2$ as a function of $x$\cite{rotter,chen3} indicate that the nature of the magnetic transition can be described as a SDW which for $x \rightarrow$ 0 appears at 148 K. As $x$ increases, the transition temperature decreases and the SDW coexists
with  superconductivitity below $x \approx$ 0.4 for the Ba$_{1-x}$K$_x$Fe$_2$As$_2$ system.
The superconducting transition temperature $T_C$ passes a maximum  of $\sim$38 K for phases with 0.4 $< x <$ 0.6, and decreases to 3.8 K for pure KFe$_2$As$_2$.
There have been several diverging reports on the magnetic and electronic
properties of single crystals of Ba$_{1-x}$K$_x$Fe$_2$As$_2$\cite{wang,Luo,ni}. Crystals of BaFe$_2$As$_2$ grown
by the self-flux method so-far were rather small but showed a metallic resistivity down to 4 K with the structural and SDW transition between 130 K to 138 K \cite{wang,Luo}.
However, significantly larger crystals obtained by the Sn-flux method
showed a semiconductor behavior with a suppression of the structural and the SDW  transition to
$\sim$85 K.  This observation has been attributed to inclusion of about 1\% Sn into the BaFe$_2$As$_2$ crystals\cite{ni}. In
Nominal Ba$_{56}$K$_{44}$Fe$_2$As$_2$ single-crystal with $T_C$'s near $\sim$27 K showed a variation of the K concentration  of up to 7\% across  the measured sample\cite{ni}.  The comparatively low $T_C$ of $\sim$ 27 K observed in these crystals  has also been associated to the inclusion of 1\% Sn.

In order to investigate these issues, especially
why the  rather large crystals grown by the Sn flux  technique exhibit  electronic and magnetic properties significantly different from
those of polycrystalline powders and crystals grown by the self flux method. We systematically modified the Sn-flux method and examined the magnetic and electronic properties of crystals obtained by
employing different growth conditions. We found, that rapid initial heating provides single-crystals with optimized magnetic and electronic properties. The influence of post-growth annealing on such crystals has also been studied.

\begin{figure}[b]
\begin{center}
\includegraphics[width=7cm]{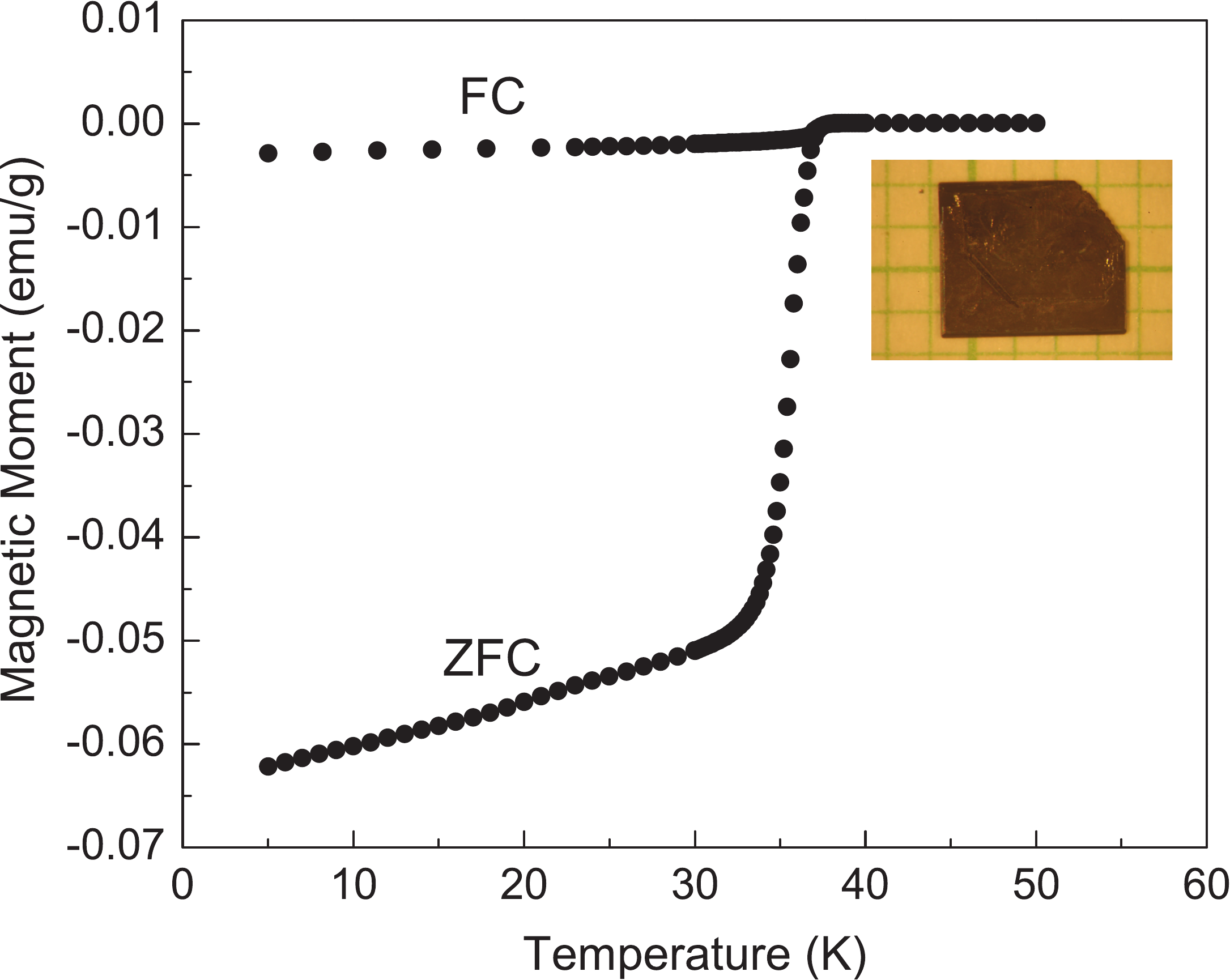}
\caption{\protect \label{Fig1}Temperature dependence of the magnetization
of a Ba$_{65}$K$_{45}$Fe$_2$As$_2$ crystal determined in a magnetic field of 10 G aligned within the $a$-$b$ plane measured in zero-field and field-cooled (fc) conditions.
Insert, picture of the $c$-axis oriented crystal with a mm scale grid paper as background.
(color online)}
\end{center}
\end{figure}

To grow single crystals of Ba$_{1-x}$K$_x$Fe$_2$As$_2$,  FeAs was used which had been obtained by  reacting stoichiometric amounts of high purity powder of Fe and As for 10 hours at 700$^{\rm o}$C in an evacuated quartz glass tube. Subsequently,  stoichiometric amounts of high purity of Ba, FeAs, and twice the stoichiometric amount of K were filled into  $\sim$2 ml alumina crucible and sealed into a quartz ampoule under $\sim$0.4 bar of high purity Ar atmosphere. In order to avoid the loss of K and formation of
K compounds below 700 $^{\rm o}$C, the quartz glass  ampoules were put directly in an oven preheated to  700 $^{\rm o}$C. The
oven was programmed to reach to 850 $^{\rm o}$C within 2 hours and stay at this temperature for another hour followed by cooling down  to 500 $^{\rm o}$C within 36 hours. Finally, the ampoule was tipped from
vertical to horizontal position at 500$^{\rm o}$C and the oven was turned off and the sample was allowed to cool down inside the oven. A large number of Ba$_{1-x}$K$_x$Fe$_2$As$_2$ crystals were obtained by this method which can be
easily separated from the solidified Sn flux (see Fig.~\ref{Fig1}) The strcuctural properties of the crystals were characterized using x-ray Laue back scattering, the chemical composition  was analyzed by microprobe analysis using  a Scanning Electron Microscope EOS TESCAN equipped with an Oxford EDX microprobe analyzing unit.

The temperature dependence of the magnetization and the electrical resistivity was determined in a Quantum-Design magnetometer (MPMS) and a physical property measurement system PPMS (Quantum Design, 6325 Lusk Boulevard, San Diego, CA.), respectively.
Several batches of single crystals of Ba$_{1-x}$K$_x$Fe$_2$As$_2$ were grown aiming at compositions  of
$x$=0.45 and $x$=0.75. Large plate like single-crystals  of size 5 $\times$ 5 $\times$ 0.15 mm$^3$ were easily separated from the Sn flux (see insert of Fig.~\ref{Fig1}). We observed traces of Sn on the surface of the crystals which can be removed mechanically or by cleaving off  the upper layers of the crystal by an adhesive tape. The microprobe analysis of the
samples revealed traces of Sn (less than 1\% ) in some of the samples while some of the cleaved samples did not show any Sn within the accuracy of the EDX system.

Laue back scattering x-ray results of crystals with composition Ba$_{55}$K$_{45}$Fe$_2$As$_2$ (according to microprobe analysis) indicated
perfect crystallographic $a$-$b$ planes (lattice parameter $a$= 0.3930 nm) but with strong disorder along the $c$ direction due to random the stacking of the $a$-$b$ sheets along
the $c$-axis ($c$= 1.313 nm).

Magnetic susceptibility measurements with the magnetic field aligned within the $a$-$b$ plane displayed in Fig.~\ref{Fig1} indicate the onset of the diamagnetic signal  at 37 K. The width of transition (10\%-90\% criterion) amounts to about 3.5 K. As is clearly seen in the zero-filed cooled (zfc) signal, the diamagnetic signal does not level off but continuous to decrease to lowest temperatures indicating some inhomogeneities in the crystals.

\begin{figure}[b]
\begin{center}
\includegraphics[width=7cm]{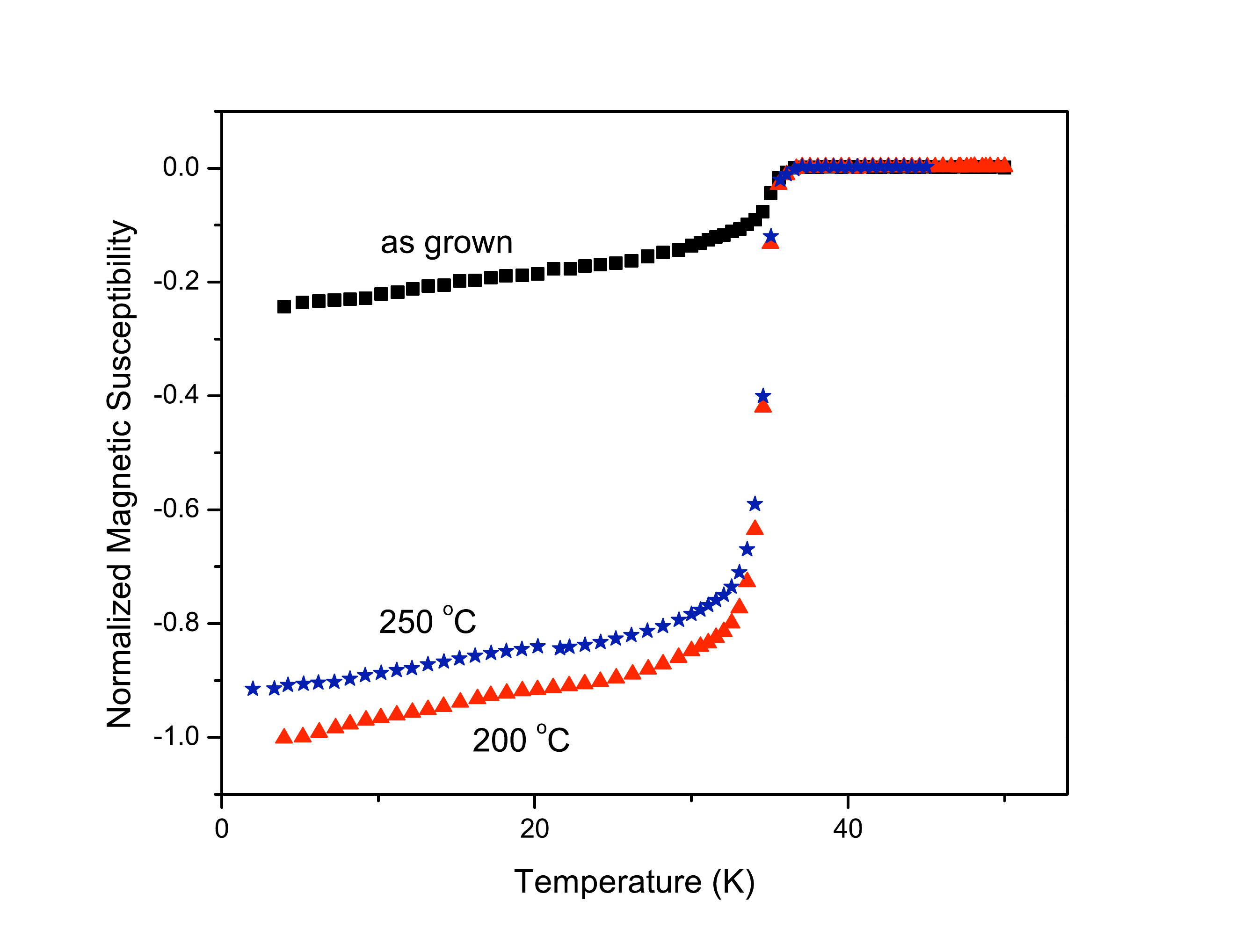}
\caption{\protect \label{Fig2}Temperature dependent of magnetization for a zero field at 10 G for as grown sample and
sample annealed at 200 $^{\rm o}$C and 250 $^{\rm o}$C.
(color online)}
\end{center}
\end{figure}
\begin{figure}[b]
\begin{center}
\includegraphics[width=7cm]{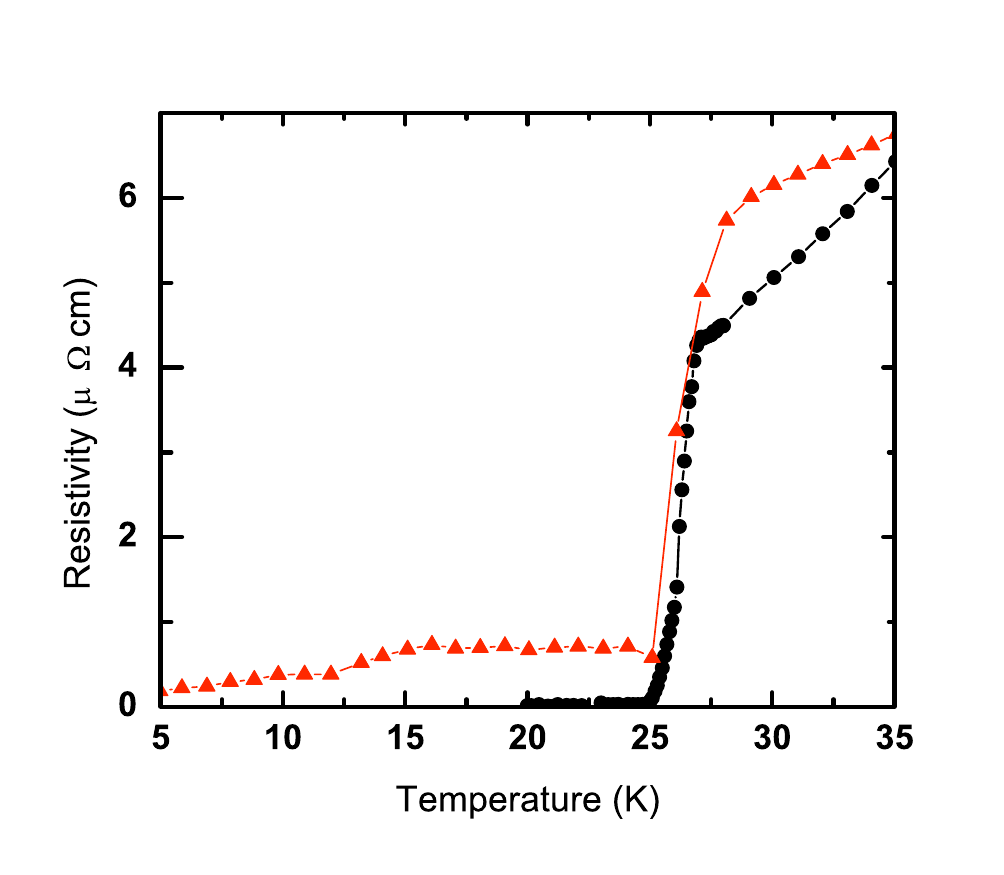}
\caption{\protect \label{Fig3}Temperature dependent of resistivity for two samples from the same batch of
Ba$_{25}$K$_{75}$Fe$_2$As$_2$.
(color online)}
\end{center}
\end{figure}

Inhomogeneities and the continuous decrease of the diamagnetic signal at low temperatures can be significantly reduced by annealing the crystals at moderate temperatures. For this a crystal was sealed in an evacuated quartz glass under He exchange gas (0.3 bar) in the tube, and after the magnetization had initially been measured, the quartz glass tube with the crystal was placed into a furnace and kept at a fixed temperature for 5 hours. After rapid quenching the magnetization of the crystals was measured again. This procedure was repeated several times with  successively increasing the annealing temperature.
After annealing at 200$^{\rm o}$C  the diamagnetic flux expulsion increased by a factor of $\sim$5 indicating increased homogeneity of the Ba/K distribution leading to a reduction  of strain effects due to an inhomogeneous arrangement of the doped K atoms.
Annealing at 250$^{\rm o}$C and  temperatures above, however, was found to reduce the diamagnetic signal Fig.~\ref{Fig2}. This result is assigned to a  loss of K and a gradual change of stoichiometry in the crystals.
To examine stoichiometry and Ba/K ordering effects we grew single-crystals aiming for a higher concentration of K namely
Ba$_{25}$K$_{75}$Fe$_2$As$_2$. Several crystals from this batch were examined by measuring their resistivity and
magnetization  as a function of temperature. All samples exhibit an onset of the diamagnetic signal due to the onset of superconductivity at about 27 K. Some crystals  however,  showed a subsequently  sharp step-shape reduction of the magnetization  at about 15 K indicating a second superconducting phase within the crystals with a $T_C$ near 15 K. An example of resistivity results obtained on two crystals with composition Ba$_{25}$K$_{75}$Fe$_2$As$_2$ is shown in  Fig. \ref{Fig3}.

\begin{figure}[b]
\includegraphics[width=7cm]{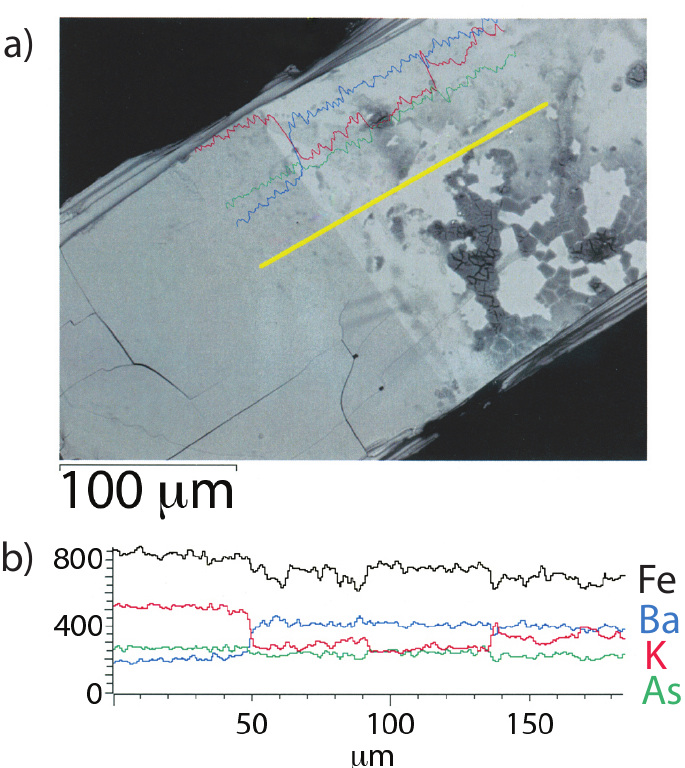}
\caption{\protect \label{Fig4}SEM image as well as a the result (traces in lower part of the figure) of a microprobe analysis scanned along a $\approx$200 $\mu$m traces of the crystal indicated by the solid (yellow) bar on the crystal. The crystal was grown in a Sn flux aiming at a composition Ba$_{25}$K$_{75}$Fe$_2$As$_2$. The speckled dark and bright areas dark area seen on the
 right hand side of the crystal are due to the adhesive used for holding the sample during resistivity measurement.
(color online)}
\end{figure}

Using SEM and electron probe micro-analysis we examined the elemental composition of one
of the crystals  which showed the second  superconducting transition with a $T_C \approx$ 15 K. The SEM pictures of the sample clearly
reveals a sharp line separating crystallites with significantly different elemental composition of Ba and K (cf. Fig. \ref{Fig4}(b))    The results of a  microanalyzer scan along a line of $\approx$200 $\mu$m of across the transition on the crystals's surface (thick line shown in the SEM
picture) is shown in Fig. \ref{Fig4}(b).
While the As and Fe content along the scan is relatively constant  a simultaneous
sharp drop in  the K  and a sharp increase of Ba content becomes evident at the phase boundary.
While in the brighter region the elemental composition  of the sample corresponds rather well
to Ba$_{0.35}$K$_{0.65}$Fe$_2$As$_2$  (lower left part of the crystal in Fig. \ref{Fig4}(a)) the composition  in the darker areas (upper right hand part of the crystal in Fig. \ref{Fig4}(a)) indicates a composition  Ba$_{0.7}$K$_{0.3}$Fe$_2$As$_2$, however with a rather large variation of the K and Fe content along the scanned line as shown in
Fig. \ref{Fig4}(b).  This compositional scatter  may be the reason why in the superconducting phase below $T_C$ the resistivity of this crystal  does not vanish (cf. Fig. \ref{Fig3}).
The sharpness of the phase boundary separating the two parts of the crystal with different elemental composition indicates that KFe$_2$As$_2$ and BaFe$_2$As$_2$ are not continuously miscible and that rather rational K to Ba ratios are assumed. This implies the question of whether or not K
atoms randomly substitute Ba atoms or whether more or less perfect K and Ba layers survive which stack in appropriate ratios.
Our preliminary
x-ray results indicate random distribution of K for Ba since no extra reflections due to
superlattice formation  due to ordering of K and Ba layers were observed.
On the other hand, the compositions found in the  two phases  are suggestive to result from a stacking of one ordered Ba and two ordered K layers (Ba$_{0.35}$K$_{0.65}$Fe$_2$As$_2$) and vice-versa of one ordered K and two ordered Ba layers (Ba$_{0.7}$K$_{0.3}$Fe$_2$As$_2$). Such a stacking of (more-or-less) perfect layers could account for the observed 1:2 (2:1)  ratios of the alkaline earths and alkaline atoms in the sharply separated phases in the investigated crystal.
We finally discuss the issue of possible Sn substitution and its influence on the superconducting properties. We studied a total of 5 crystals showing very similar characteristics in their superconducting properties as the crystal described in detail before. These investigations gave no indication that
Sn impurities in quantitative amounts are present which could play a decisive  role for the variation of the superconducting properties of these compounds. We conclude that  rather the determining factor is the variation of the K and Ba content and possibly the stacking of more-or-less non-substituted Ba and K layers throughout the crystal is the most
important factor in determining the electronic and magnetic properties. A careful examination of crystal growth conditions and the microscopic distribution of Ba and K atoms seems to be necessary to optimize the superconducting properties of these phases.

\acknowledgments Financial support for this work was partially provided by the Natural Sciences and Engineering Research Council of Canada. We thank C. Kamella, E. Br\"ucher and G. Siegle for experimental assistance.

\end{document}